\documentclass[aps,pra,twocolumn,tightenlines,longbibliography,superscriptaddress,doublespace,nofootinbib]{revtex4-2} 

\usepackage{graphicx}
\usepackage{setspace,appendix,physics,wasysym,xcolor,xparse}
\usepackage{color}
\usepackage{appendix} 
\usepackage[english]{babel}
\usepackage{amsmath,amssymb,amsfonts,mathtools}
\usepackage{mathtools}
\definecolor{brickred}{rgb}{0.8, 0.25, 0.33}
\definecolor{celestialblue}{rgb}{0.29, 0.59, 0.82}
\definecolor{cornflowerblue}{rgb}{0.39, 0.58, 0.93}
\definecolor{denim}{rgb}{0.08, 0.38, 0.74}
\definecolor{armygreen}{rgb}{0.29, 0.33, 0.13}
\definecolor{cardinal}{rgb}{0.77, 0.12, 0.23}
\definecolor{carnelian}{rgb}{0.7, 0.11, 0.11}
\definecolor{armygreen}{rgb}{0.29, 0.33, 0.13}

\usepackage[colorlinks=true, urlcolor=black, citecolor=blue, linkcolor=red, citebordercolor={1 0 0}, linkbordercolor={0 0 1}]{hyperref}

\usepackage{verbatim,multirow}
\usepackage[T1]{fontenc}
\usepackage{lmodern}
\usepackage{pxfonts}
\usepackage{orcidlink}

\begin{document}
\title{Quantum random number generation from the continuous variable payload for the SPOQC mission}
\author{Vinod N. Rao\,\orcidlink{0000-0003-0364-1834}}
\email{vinod.rao@york.ac.uk}
\affiliation{School of Physics, Engineering \& Technology and York Centre for Quantum Technologies, University of York, YO10 5FT York, U.K.}
\affiliation{Quantum Communications Hub, University of York, U.K.}
\author{Killian Murphy\,\orcidlink{0000-0001-5665-8248}}
\affiliation{Quantum Communications Hub, University of York, U.K.}
\author{Fadi Ahwal\,\orcidlink{0000-0002-5512-0275}}
\affiliation{School of Physics, Engineering \& Technology and York Centre for Quantum Technologies, University of York, YO10 5FT York, U.K.}
\author{Emma Tien Hwai Medlock\,\orcidlink{0000-0001-5852-5510}}
\affiliation{School of Physics, Engineering \& Technology and York Centre for Quantum Technologies, University of York, YO10 5FT York, U.K.}
\author{Timothy P. Spiller\,\orcidlink{0000-0003-1083-2604}}
\affiliation{School of Physics, Engineering \& Technology and York Centre for Quantum Technologies, University of York, YO10 5FT York, U.K.}
\affiliation{Quantum Communications Hub, University of York, U.K.}
\author{Rupesh Kumar\,\orcidlink{0000-0002-6813-6401x}}
\email{rupesh.kumar@york.ac.uk}
\affiliation{School of Physics, Engineering \& Technology and York Centre for Quantum Technologies, University of York, YO10 5FT York, U.K.}
\affiliation{Quantum Communications Hub, University of York, U.K.}

\begin{abstract}
The necessity of random numbers for various tasks, from simulation to cryptography, is crucial and immense. Here we demonstrate CV-QRNG using the CV payload of the SPOQC mission. The homodyne setup for QRNG uses the laser from the payload, in addition to potentially being used as detector in the case of an uplink scenario. Here we quantify the extractable secure randomness from the QRNG setup, that involves homodyne measurement of the vacuum states. The extracted randomness is tested against NIST test suite in addition to formally upper bounding the min-entropy. With the raw key length being $\approx1$ Mb in a given satellite pass, we get a total length of $\approx19.5$ Kb of certified random numbers from the 12-bit ADC. 
\end{abstract}

\maketitle

\section{Introduction \label{sec:intro}}

The requirement of random numbers ranges from cryptography and finance, to various experiments in physics \cite{herrero2017quantum}. There are many physical sources such as cosmic rays, white noise, radioactivity, etc, through which one can extract randomness. Another important approach to generate random numbers is by using computational algorithms. The latter has been of use since the advent of digital age, and it is currently the most pertinent source of randomness. However, all of these sources offer randomness based on an underlying unpredictability from classical or mathematical complexity class. Contrary to this is the quantum random number generators (QRNG) that offer information-theoretic secure randomness \cite{colbeck2011private, huang2020practical}.

The idea of extracting randomness from a quantum system is based on the aspect of utilizing the inherent probabilistic nature of quantum mechanics. There are various resources in quantum mechanics, such as entanglement, detection of vacuum fluctuations and quantum coherence \cite{ma2016quantum, stefanov2000optical}, which have a specific kind of uncertainty. The most basic idea of QRNG is that of measuring a pure state with projective measurement on a non-orthogonal basis. The interpretation of quantum mechanics then states that the possible outcomes are genuinely random with probabilities given by the Born rule \cite{busch2003quantum}.

Quantum information using continuous variables (CV) deals with the preparation and measurement of states in an infinite dimensional Hilbert space \cite{braunstein2005quantum, weedbrook2012gaussian}. Thus, extracting randomness from CV states is of great interest, focused particularly on the Gaussian states such as extraction of randomness from the vacuum states \cite{zheng20196, bruynsteen2023100, zhou2019practical} and squeezed states \cite{cheng2024quantum, cheng2024semi}. Various other resources can be used to extract randomness, such as non-Gaussian states \cite{ra2020non, walschaers2021non}, quantum entanglement \cite{pironio2010random, liu2018device}, photonic statistics \cite{applegate2015efficient, aungskunsiri2023quantum}, spontaneous emission process \cite{guo202140}, modes of orbital angular momentum \cite{zahidy2022quantum, wang2022quantum}, photonic waveguides \cite{strydom2024quantum}, phase noise of the laser \cite{qi2010high, abellan2014ultra} and others \cite{martinez2018advanced, mannalatha2023comprehensive}. 

CV-QRNGs based on homodyne measurement of vacuum states have been proposed and implemented \cite{shen2010practical, xu2012ultrafast, shi2016random, zheng20196, marangon2017source, guo2018enhancing}. These involve utilizing the randomness from the quantum fluctuations of the vacuum state, and quantifying them by suitably removing the classical correlations from within \cite{gabriel2010generator, shrivastava2025randomness}. These classical correlations may come from various contributing factors such as: electronic noise, imbalance in homodyne setup, detector inefficiencies, etc \cite{zhang20181, ferreira2021homodyne, li2023practical}. The classical noise contributions from the above mentioned sources could be due to different constituent elements of the quantum system's and measurement system's interaction with the environment. The underlying security comes from the fact that the information leakage from using vacuum states is zero. Ideally, the intrinsic randomness in measuring the quadratures of a vacuum mode stems from the lowest energy state (vacuum), which cannot be correlated to any other state \cite{gabriel2010generator}. However, since one can never quantify the correlations due to noise, if there are any, a non-zero leakage is attributed due to a non vacuum state \cite{guo2018enhancing}. Thus, the extraction of randomness from the output is an important process, which provides a specific output length of random numbers. in this work, we employ Toeplitz matrix extractors, but various other extractors can be used as well. 

As part of the payload developed for demonstrating space-to-ground CV Quantum Key Distribution (QKD), we have incorporated a homodyne based CV-QRNG in the Satellite Platform for Optical Quantum Communication (SPOQC) mission - an in-orbit demonstrator for quantum communication by the UK Quantum Communications Hub \cite{SPOQC, IQN}. The results provided here are from the engineering model of the CV payload, which is identical to the flight model. The payload contains a homodyne detector followed by a 12-bit Analog-to-Digital Converter (ADC). Additionally, a Single Board Computer (SBC) is also present where the randomness extraction happens. 

The primary motivation and outcome of the work is to show the possibility of random number generation in space for various applications. This is an important result as the payload is having various limitations such as size, weight and power (SWaP) constraints. Here we show that we can generate a small, but secure random numbers in SPOQC CV payload which is intended to use for QKD post processing tasks such as basis selection and error correction. Additionally, this also opens up the possibility of using these random numbers for modulating the signal. This manuscript is organised as follows. After a brief introduction in Sec. \ref{sec:intro}, we provide the main results in Sec. \ref{sec:cvqrng}. Finally, the conclusions are given in Sec. \ref{sec:conc}. 

\section{CV-QRNG \label{sec:cvqrng}}

The CV-QRNG based on vacuum noise detection involves a shot noise limited homodyne detector, as given in Fig 1. The shot noise quadrature measurement contains total noise, including the quantum as well as the classical part. Here we estimate the extractable amount of random numbers by removing the contribution from the classical noise in Sec. \ref{sec:form}. Followed by the design of the QRNG module in the CV payload in Sec. \ref{sec:qrng}. Finally the results are presented in Sec. \ref{sec:res}.

\begin{figure}[h]
\centering
\includegraphics[scale=0.4]{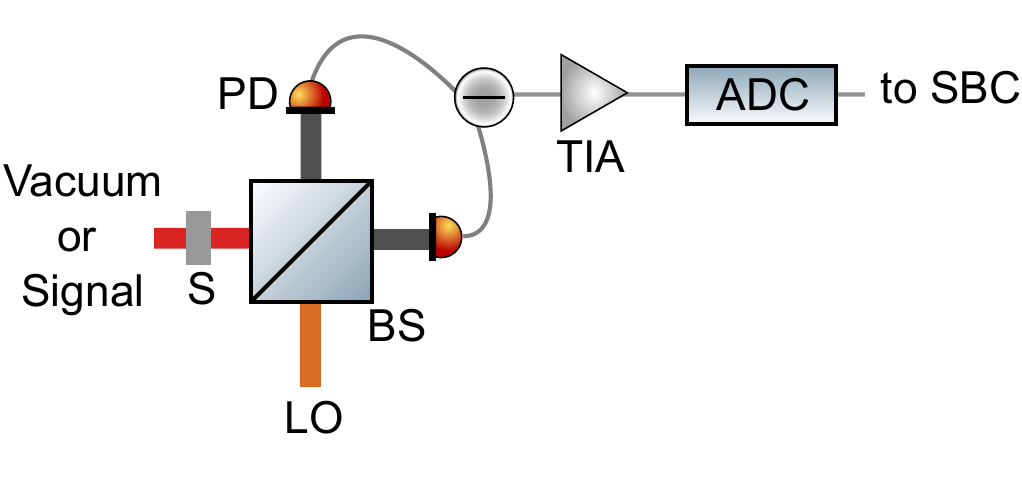}
\caption{Homodyne measurement of two signals. LO - Local Oscillator; BS - Beam-Splitter; PD - Photo-Detector; TIA - Trans-Impedance Amplifier; ADC - Analog-to-Digital Converter; PC - Computer.}
\vspace{0.5cm}
\label{fig:homo}
\end{figure}

\subsection{Formalism \label{sec:form}}

Here we provide the brief details of the formalism to estimate the various loss and noise factors in a general homodyne setup. Taking the contribution of the classical noise to be independent, we can write the Shannon entropy of the quantum part $H(X)_{q}$ as:

\begin{equation}
H(X)_{q} = H(X)_{total} - [H(X)_{c(i)}],
\label{eq:entropy}
\end{equation}

where $H(X) = -\sum p_i ~ \text{log}_2(p_i)$, where $p_i$ is the probability distribution of variable X; $c(i)$ corresponds to various contributing factors to the classical noise. However, there is a limitation in the aspect of finding the true uncertainty or randomness from the quantum part. The formalism to estimate the entropy from classical sources is considered, but not quantum part. It is assumed that Eve can obtain full information of the entropy from classical sources and some information from the quantum part of the entropy as well. We determine the total knowledge of Eve in this case, using the conditional-min entropy as follows.

Considering the conditional min-entropy that quantifies Eve's information, the length of extractable randomness is given by 
\begin{equation}
k \ge l H_{\text{min}}(X \mid E) - \log_2 \left(\frac{1}{2 \epsilon_{\text{hash}}^2}\right),
\end{equation}
where $H_{\text{min}}(X \mid E)$ is the conditional min-entropy of random variable $X$ with the quantum side-information $E$ leaked to Eve, $l$ being the number of samples and $\epsilon_{\text{hash}}$ is the security parameter which depends on the distance between a random string and the randomness extractor.

Given that the quadrature operator measurement made on the vacuum mode is $\hat{X}_a$, $g$ is the gain factor and $\hat{N}$ is the noise operator of all the noise sources, the measurement can be written as 
\begin{equation}
\hat{q} = g (\hat{X}_a + \hat{N}). 
\end{equation}

Any measurement from the homodyne detector setting outputs a value which would have a probability distribution. Let the random variable to which this distribution belongs be $X$. Assuming that the input to the homodyne measurement is a thermal state with the mean photon number $n$, then the probability density distribution is given by
\begin{equation}
p_X(x) = G(x;0,g^2 (1+2n)), 
\end{equation}
where $x$ is the output of variable $X$, $g$ is the gain from the measurement and
\begin{equation}
G(x;\mu, \nu^2) = \frac{1}{\sqrt{2\pi \nu^2}} e^\frac{-(x-\mu)^2}{2\nu^2} 
\end{equation}
is the Gaussian distribution in variable $X$, with mean $\mu$ and variance $\nu^2$.

In the present case, the variable $X$ is discretized by the ADC of range $R$ and bin size $\Delta x$. Thus, the continuous variable $X$ is replaced by its discretized variable $\overline{X}$, which has values $j = 1,2,\dots d$ and the distribution 
\begin{equation}
p_{\overline{X}}(j) = \int_J p_X(x) dx,
\end{equation}
with $J$ being the $d$ intervals that discretize the outcome of homodyne detection. The correlation between $\overline{X}$ and $E$ is 
\begin{equation}
\rho_{\overline{X}E} = \sum_j p_{\overline{X}}(j) \ket{j}\bra{j} \otimes \rho_E^{(j)},
\end{equation}
with $$\rho_E^{(j)} = \frac{1}{p_{\overline{X}}(j)}\int_J p_X(x)\rho_E^x dx .$$
This is a general description of Eve's state, where her output is correlated to the QRNG output via the input thermal state $p_X(x)$. The correlations between $\overline{X}$ and $E$ are important as they provide the formalism to estimate Eve's correlation with the output. Let the measurement outcomes $q_k$ from the ADC on an arbitrary input state $\rho_A$ 
be represented by probability $p(q_k) = \text{Tr} [\rho_A Q^k_{\delta q}]$, which are stored in a classical register $Q^k_{\delta q}$ that records the output from the homodyne setting. This is just the measurement operator from which output is obtained. The classical register here represents the output across the ADC in our case. Thus, given the output data set and the assumed input state being thermal, we find the probability of occurrence of each output as given in Fig. \ref{fig:G_SN} and Fig. \ref{fig:G_EN}. 

The variance of the Gaussian distribution $\nu^2$ corresponds to the shot noise variance of the detectors in the homodyne measurement, which in turn depends on the shape of the temporal mode of the detectors. Ideally, Eve would not have all the information from the source as the input state is pure, and the measurement being projective. However, this is not the case as the verification of the input state being uncorrelated could never be done in a device dependent scenario like ours. The various noises detected would thus have correlations between input state and output state and thus require a quantification of entropy of output given information leakage to Eve. This is done using the conditional min entropy, as the Shannon entropy usually quantifies the average unpredictability of a probability distribution whereas the min entropy estimates the difficulty in predicting the most probable outcome. The general min-entropy of the output is given by 
\begin{equation}
H_{\text{min}}(X)_{\rho_{\overline{X}}} = \frac{1}{2}{\log_2 (2\pi \nu^2)}.
\label{eq:minent}
\end{equation}

As we do not trust the source of randomness, let us assume that the arbitrary input state $\rho_A$ could be correlated with the state of Eve. This corresponds to $\rho_A$ being mixed and $\rho_A = \text{Tr}_E[\rho_{AE}]$, with $\rho_{AE}$ being the joint state. After the projective measurement on system $A$ with outcome $k$, Eve’s state is projected to $\rho_E^k$. The total collective state is now a classical-quantum state, 
\begin{equation}
\rho_{QE} = \sum_k p(q_k) \ket{k}_A\bra{k} \otimes \rho_E^k,
\label{eq:jost}
\end{equation} 
with $Q$ representing the detected state from homodyne measurement, with probability distribution $p(q_k)$. This joint state could be used to quantify the conditional min-entropy from the homodyne measurement of the thermal state.

Another way to quantify the extractable randomness from such classical-quantum states is using the single-shot entropy estimation \cite{anshu2019minimax, wang2025one}. These are shown to work for both finite and infinite dimensional reduced quantum states. The maximum amount of secure extractable randomness, from such single-shot estimation of output from the classical register $Q^k_{\delta q}$, is given by 
\begin{equation}
r_{\text{sec}}^\epsilon (Q^k_{\delta q} \mid E) = H_{\text{min}} (Q^k_{\delta q} \mid E) - 2 \text{log}_2 \frac{1}{\epsilon},
\end{equation}
where $\epsilon$ is the security parameter and $H_{\text{min}} (Q^k_{\delta q} \mid E)$ is the conditional min-entropy. This is different from the general case of min-entropy given in Eq. (\ref{eq:minent}) in that the conditional case estimates the entropy of the system conditioned on Eve's state $E$. Thus, this quantifies the information leaked to Eve based on the type of attack she would perform and her potential state. Therefore, the protocol of extracting randomness from homodyne measurement of vacuum states is then said to be $\epsilon$-secure, which means that the probability of distinguishing the output from a truly uniform independent distribution is smaller than $\frac{1}{2} (1+\epsilon)$. The conditional min-entropy is defined as, 
\begin{equation}
H_{\text{min}} (Q^k_{\delta q} \mid E) = -\text{log}_2 ~ \underset{\{E_k\}}{\text{max}} \sum_k ~ p(q_k) \text{Tr}[E_k \rho_E^k],
\end{equation}
with $\{E_k\}$ being the POVM on system $E$. The quantity $p_{\text{guess}} (\{E_k\}) = \sum_k p(q_k) \text{Tr}[E_k \rho_E^k]$ is the average probability for the adversary Eve to correctly guess the index $k$ using a measurement strategy $\{E_k\}$. Here, $k$ represents the respective POVM element with $\sum_k E_k = I$.

The maximization of the POVM $\{E_k\}$ corresponds to finding the best measurement strategy Eve might apply to guess the index $k$ of the post measurement state $\rho_{QE}$ in Eq. (\ref{eq:jost}). The amount of secure randomness is then the smallest conditional min-entropy for states $\rho_{QE}$ consistent with the input state $\rho_A$. If the state $\rho_A$ is pure, this implies that $A$ and $E$ are independent, $\rho_{AE} = \rho_{A} \otimes \rho_{E}$, in which case the conditional min-entropy reduces to the classical unconditional min-entropy
\begin{equation}
H_{\text{min}} (Q^k_{\delta q}) = -\text{log}_2 ~ \underset{k}{\text{max}} ~ \{p(q_k)\}. 
\end{equation}

Here, Eve’s best strategy is to guess the most likely index $k$ every time. For any state, $H_{\text{min}} (Q^k_{\delta q}) \ge H_{\text{min}} (Q^k_{\delta q} \mid E)$ and the difference can be seen as the amount of side information accessible to Eve. To compute the exact value of $H_{\text{min}} (Q^k_{\delta q} \mid E)$, one needs to know $\rho_{QE}$. Since we do not have access to $E$, we would need to perform a complete tomography of the input state $\rho_{A}$ to find all compatible states $\rho_{QE}$. This is tedious for an infinite-dimensional system like ours, where the input state is assumed to be thermal. Instead, one can bound $H_{\text{min}} (Q^k_{\delta q} \mid E)$ by the max-entropy of the conjugate quadrature $H_{\text{max}} (P_{\delta p})$ using the entropic uncertainty relation for QRNG as \cite{marangon2017source},
\begin{equation}
H_{\text{min}} (Q^k_{\delta q} \mid E) + H_{\text{max}} (P_{\delta p}) \ge -\text{log}_2 ~ c(\delta c, \delta p), 
\end{equation}
where the overlap constant $c(\delta c, \delta p)$ signifies the upper bound on the conditional min-entropy and the max-entropy with the probability distribution $p(p_k)$ is defined as 
\begin{equation}
H_{\text{max}} (P_{\delta p}) = 2\text{log}_2 \sum_k p(p_k). 
\end{equation}

Given the density operator of the Eve being $\gamma_E$, the conditional min-entropy thus given by
\begin{align}
H_{\text{min}} (\overline{X} \mid E) &\ge -\text{log}_2 \big[\underset{j}{\text{sup}}~(p_{\overline{X}}(j) \mid\mid \gamma_E^{-1/2} \rho_j \gamma_E^{1/2} \mid\mid_\infty)\big] \nonumber \\
&= -\text{log}_2 (\sqrt{n} + \sqrt{n+1})^2 -\text{log}_2 \left[ \text{erf}\left(\frac{\Delta x}{2g^\prime}\right) \right],
\label{eq:c-min}
\end{align}
with $\Delta x$ representing the bin size of an ideal ADC and $g^\prime$ being the new gain factor.

This is the conditional min-entropy with an ideal ADC, as we have not considered the ADC digitisation error until now. However, if we assume the ADC digitisation has the binning of $M$ possible outcomes, then the reduction in ADC range is at most $\log(M)$ bits. Thus the conditional min-entropy including ADC error is given by 

\begin{equation}
H_{\text{min}} (\overline{X} \mid E, N) = H_{\text{min}} (\overline{X} \mid E) -\log[~\text{sup} \mid J_f \mid],
\label{eq:ADCerror}
\end{equation}
where $J_f$ denoting any output $f$ that is mapped from the $j$ true values due to digitisation.

Based on the expression in Eq. $\eqref{eq:ADCerror}$, the conditional min-entropy can be numerically calculated. The min-entropy depends on the variance of the quantum and the classical noise, the range $R$ of the ADC and the total number of bins $M$. For a given range of the ADC, one can optimise the conditional min-entropy over the possible outcomes.

\subsection{QRNG module in CV payload \label{sec:qrng}}

\begin{figure}[h]
\centering
\includegraphics[scale=0.5]{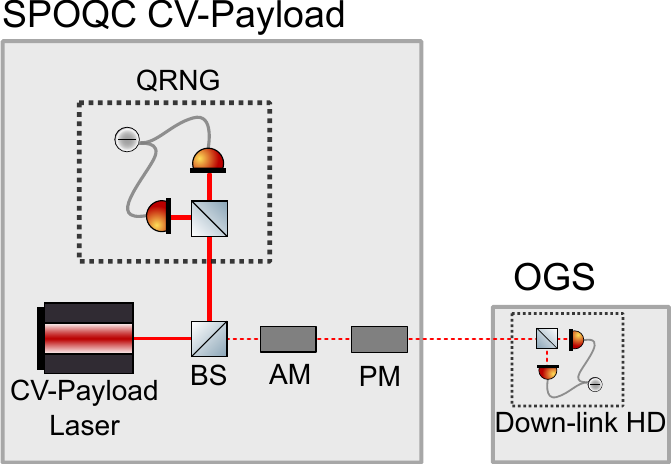}
\caption{An abstract block diagram of the CV payload and receiver for the SPOQC mission. Gaussian modulated coherent states are created using amplitude and phase modulator (AM \& PM) and sent to the optical ground station (OGS). A beam-splitter (BS) is used to send some part of signal and/or LO to QRNG receiver module. HD is the homodyne detector.} 
\vspace{0.5cm}
\label{fig:payload}
\end{figure}

In the following, we provide the details of the setup of the QRNG in the payload. An abstract block diagram of the payload is shown in Fig. \ref{fig:payload}. Though the primary objective of the mission is to demonstrate downlink based CV-QKD from Low Earth Orbit (LEO), we have incorporated a CV-QRNG setup in the payload. The optical components of the CV-QRNG, such as laser, beam splitter and detectors are part of the CV-QKD payload in the SPOQC satellite. The same setup is used for detecting the generated CV-QKD signals for calibration purposes. The CV payload-transmitter involves a $2~\text{ns}$ pulsed laser of wavelength $1550.12~\text{nm}$, operating at $2~\text{MHz}$ repetition rate. The laser is optically modulated and attenuated to a level suitable for CV-QKD demonstration. Another part of the laser, which serves as the local oscillator (LO), is sent to a shot noise limited homodyne detector which acts as the source for the QRNG. 

The sequence of operation is initiated by setting the homodyne detection unit in QRNG mode which blocks the monitoring photon detector current from the  CV-QKD signal reaching the QRNG electronic board, with the use of a digitally controlled relay switch. The electronic noise quadrature data is taken prior to every QRNG session to accommodate temperature related drift in electronic noise. Temperature drift cause the laser operating wavelength to drift, but it has no influence in the shot noise level. The other classical noise from imbalance and detector efficiencies is calibrated in advance (in the lab prior to the satellite launch). The common mode rejection ratio of the homodyne detector is >50dB and influence of laser intensity fluctuation is negligible and attributed to the input noise being thermal (Eq.4).  The laser is then switched on and the shot noise quadrature data is acquired. The QRNG optical part consists of a radiation hardened 50/50 fiber beam splitter and two vacuum compatible PIN photodiodes. The photocurrents are subtracted and amplified by low noise amplifiers (AMPTEK A250F) and digitised by a 12-bit ADC(Analog Device AD9220). The ADC outputs are acquired by the single board computer at the rate of $500~\text{kHz}$ - the sampling rate is hardware limited. The optical fibres associated with the beam splitters and photodiodes are radiation hardened to mitigate any degradation in transmission due to space radiation. These components have also been verified through vacuum and thermal cycling tests to ensure reliable operation under space conditions. Additionally, the single board computer employs redundant data storage for the operating system boot image and has similarly undergone vacuum and thermal testing to verify its performance in the space environment.

The details of the experimental method are as follows. For the extraction of potential randomness from vacuum states, the following datasets from three different configurations are taken.
\begin{enumerate}
\item Varying the optical power of the LO, to determine $H(X)_{c(1)}$ -- noise due to imperfect homodyning.
\item LO off while all the electrical units are on, to determine $H(X)_{c(2)}$ -- electronic noise.
\item LO and electrical units are on, to determine $H(X)_{total}$ -- shot noise. 
\end{enumerate}

However, amongst all of the sources that contribute to classical entropy, only the $H(X)_{c(2)}$ is observed to be dynamic and other noise sources are taken to be constant at a given temperature and setting. This assumption is based on the fact that the variation in beam-splitter reflectivity, detector inefficiencies, etc., are almost negligible (Fig. \ref{fig:homo}). However, various environmental factors such as temperature could potentially affect the performance of these units. Thus, we are restricting ourselves to operate in QRNG mode only within a range of temperature and pressure, where they have been characterised ( 1 bar Pressure at $25^\circ$ Celsius). The environmental conditions during the flight will be micro-bar pressure and temperature in the range $0^\circ- 25^\circ$ Celsius. Therefore, given $H(X)_{c(1)}$ remains constant, only two configurations would be required to estimate $H(X)_{total}$ and $H(X)_{c(2)}$ in every QRNG run. However, characterising detector efficiencies for the specific wavelength and estimating the ADC loss by suitably choosing the bin size \& range after discretisation are equally important as well. We assume that the detector efficiency is flat within the wavelength range around 1550nm and the bin size is fixed.

\subsection{Results \label{sec:res}}

\begin{figure}[h]
\centering
\includegraphics[scale=0.5]{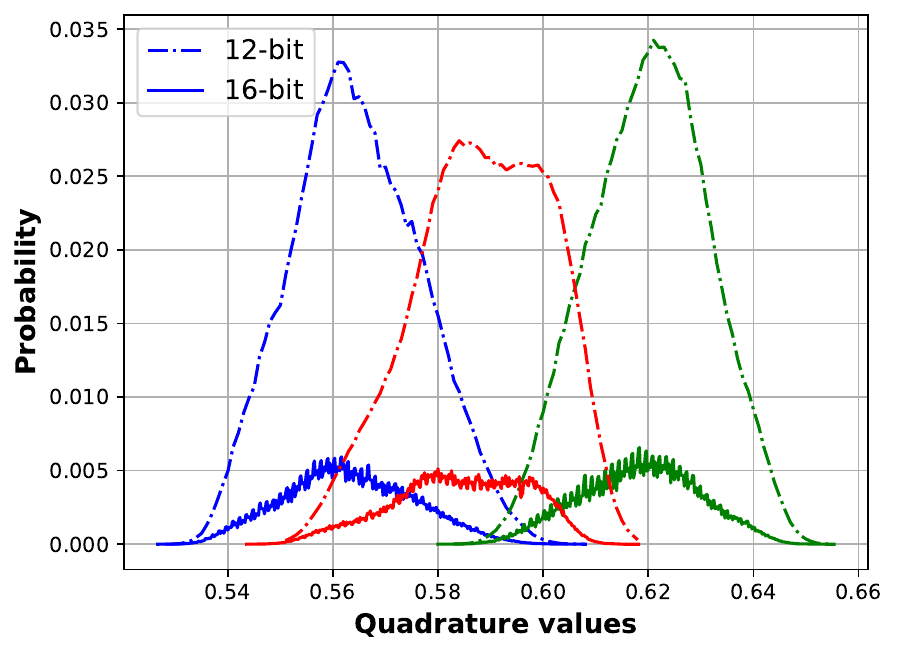}
\caption{Probability distributions of the shot noise. The colours represent three data sets.}
\vspace{0.5cm}
\label{fig:G_SN}
\end{figure}

\begin{figure}[h]
\centering
\includegraphics[scale=0.5]{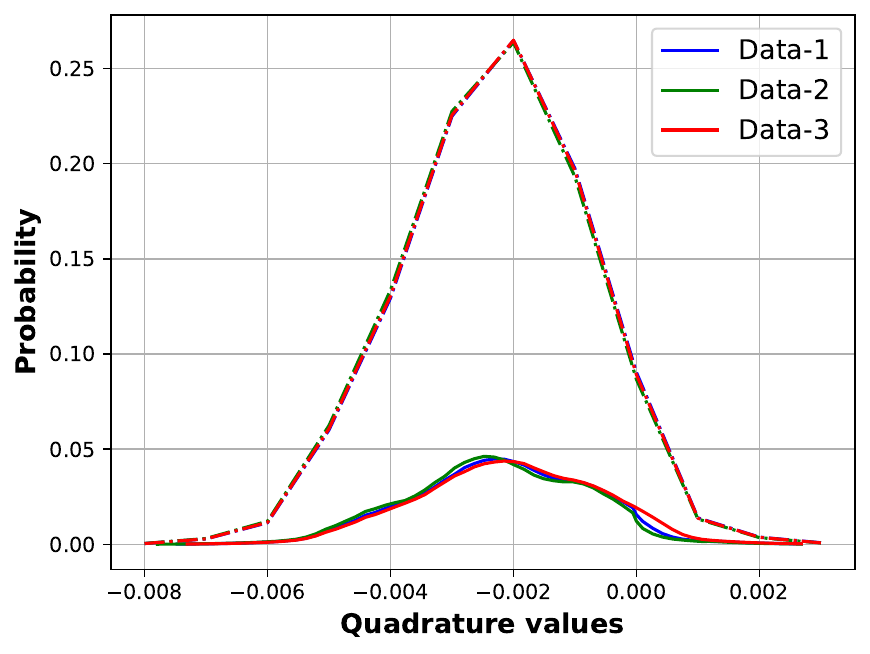}
\caption{Probability distributions of the electronic noise. The dashed (dotted) line corresponds to output from 12-bit (16-bit) ADC.}
\vspace{0.5cm}
\label{fig:G_EN}
\end{figure}

Shown in Figs. \ref{fig:G_SN} \& \ref{fig:G_EN} are the probability distributions of the shot noise and electronic noise, respectively, of three instances of the QNRG run. The probabilities of outcomes are plotted against the respective homodyne output values. Two ADC ranges are considered, 12-bit and 16-bit. The offset from zero mean in Fig. \ref{fig:G_SN} is due to the balancing. Coherent detection involves balancing the homodyne detection system. This process equates the photocurrents from both photodiodes so that their difference, which is referenced against the LO port, yields a zero mean value. The probability for the raw data is plotted here and the drift can be brought back to zero digitally in the post-processing.

From these respective probability distributions, we find the averaged entropy values as below. In Table \ref{tab:entropies}, the respective entropy values are given, estimated from the three trial cases. 
\begin{enumerate}
\item For the 16 bit ADC, 
\begin{enumerate}
\item $H(X)_{total} \approx 8.118$.
\item $H(X)_{c(2)} \approx 5.159$.
\end{enumerate}
\item For the 12 bit ADC, we find that
\begin{enumerate}
\item $H(X)_{total} \approx 4.520$.
\item $H(X)_{c(2)} \approx 2.522$.
\end{enumerate}
\end{enumerate}

\begin{table}[h]
\centering
\begin{tabular}{|c|c|c|c|c|c|c|c|}
\hline
 & \multirow{2}{*}{Case \#} & \multicolumn{3}{c|}{$H(X)$} & \multicolumn{3}{c|}{$H_{\text{min}}(X)$} \\
\cline{3-8}
 & & $\cdot_{total}$ & $\cdot_{c(2)}$ & $\cdot_{q}$ & $\cdot_{total}$ & $\cdot_{c(2)}$ & $\cdot_{q}$ \\
\hline%\cline{1-4}
\hline
\multirow{3}{*}{16-bit ADC} & (1) & 8.021 & 5.122 & 1.999 & 0.605 & 0.381 & 0.124 \\
\cline{2-8}
 & (2) & 8.231 & 5.232 & 1.999 & 0.604 & 0.378 & 0.126 \\
\cline{2-8}
 & (3) & 8.103 & 5.124 & 1.979 & 0.639 & 0.416 & 0.123 \\
\hline
\hline
\multirow{3}{*}{12-bit ADC} & (1) & 4.518 & 2.520 & 1.458 & 0.912 & 0.518 & 0.294 \\
\cline{2-8}
 & (2) & 4.519 & 2.521 & 1.448 & 0.937 & 0.527 & 0.310 \\
\cline{2-8}
 & (3) & 4.524 & 2.525 & 1.459 & 0.888 & 0.506 & 0.282 \\
\hline
\end{tabular}
\caption{The respective entropic values are listed.}
\label{tab:entropies}
\end{table}

\begin{figure}[h]
\centering
\includegraphics[scale=0.45]{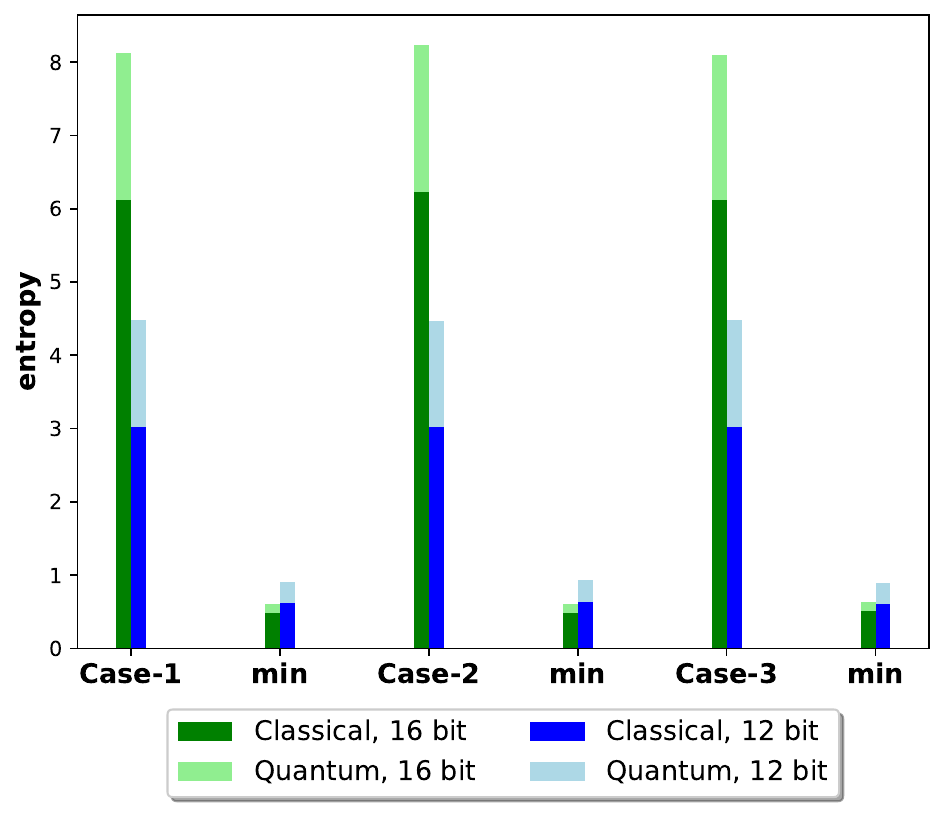}
\caption{Respective entropy values.}
\vspace{0.5cm} 
\label{fig:entropy}
\end{figure}

\begin{figure}[h]
\centering
\includegraphics[scale=0.5]{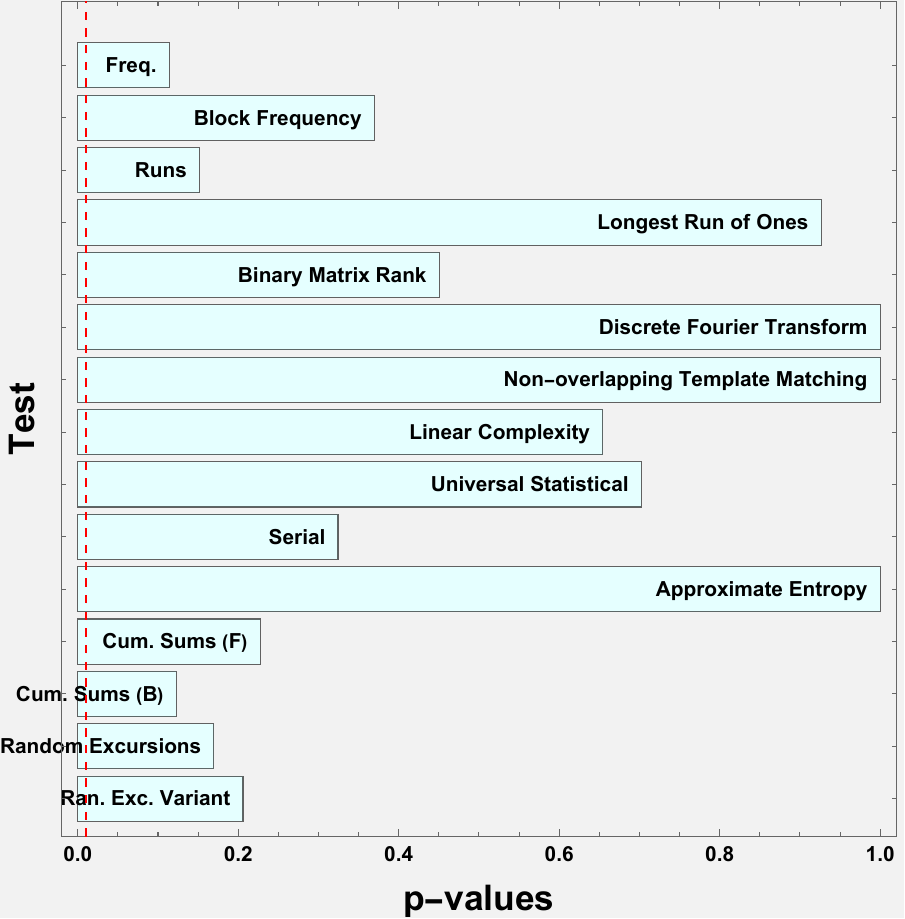}
\caption{Results from NIST test, with the p-values. The minimum p-value for the tests is given in red dashed line.}
\vspace{0.5cm} 
\label{fig:nist}
\end{figure}

Below are the figures representing the respective (min-)entropy values in Fig. \ref{fig:entropy} and the output of the NIST test suite in Fig. \ref{fig:nist}. In Fig. \ref{fig:nist}, the minimum p-value, that represents the success of the test, is given as a red dashed line.

\begin{figure}[h]
\centering
\includegraphics[scale=0.5]{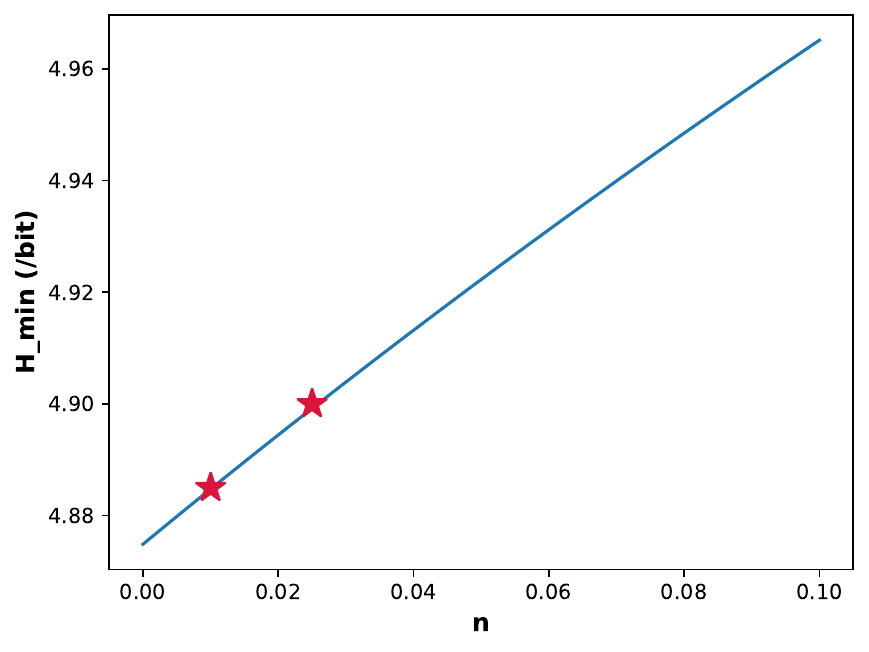}
\caption{$\text{H}_{\text{min}}$ as a function of mean photon number $n$. The stars here represent the $H_{\text{min}}$ value for 12- and 16-bit ADC outputs (smaller value for 12-bit).}
\label{fig:Hmin}
\end{figure}

\begin{figure}[h]
\centering
\includegraphics[scale=0.5]{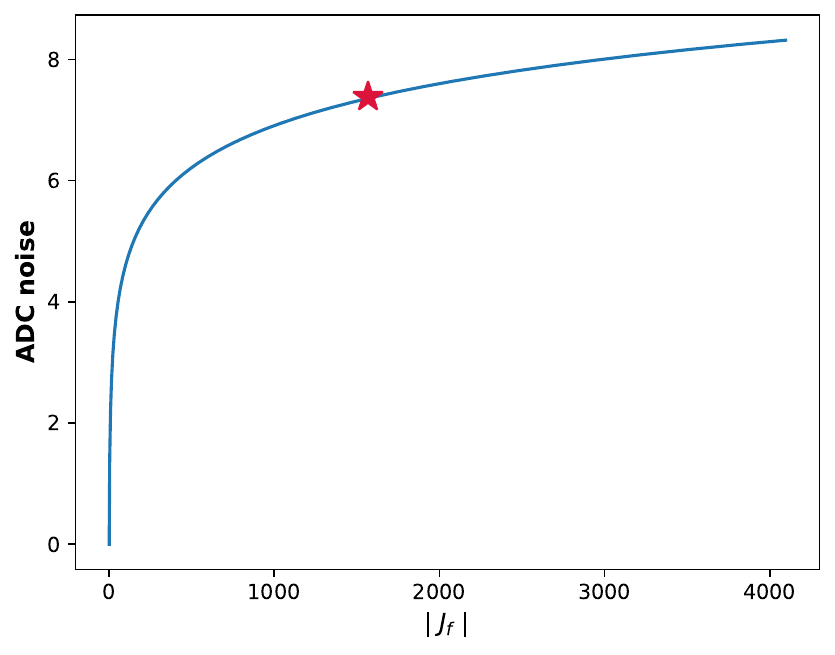}
\caption{ADC noise as a function of cardinality. The star here represents the ADC error for a 12-bit ADC on the SBC, which has a resolution of 4095 points. But the distinct values in the current setup number is under 2000.}
\label{fig:ADC}
\end{figure}

Below plotted in Fig. \ref{fig:Hmin} and Fig. \ref{fig:ADC} are the graphs corresponding to Eqs. (\ref{eq:c-min}) and (\ref{eq:ADCerror}), wherein $H_{\text{min}}$ and the ADC error are shown respectively. We note that the results presented here correspond to the operation of the CV-QRNG module at a temperature of $25~^{\circ}\text{C}$ and a pressure of 1 atm. The module has also been tested across a temperature range of $-20~^{\circ}\text{C}$ to $+55~^{\circ}\text{C}$ and under pressure conditions varying from 1 atm down to $10^{-6}$ atm.

\section{Discussions and conclusions \label{sec:conc}}

A CV-QRNG based on homodyne measurement of vacuum fluctuations is studied and results are reported. The CV payload for the SPOQC mission contains a laser that outputs a Gaussian modulated signal \& LO for downlink based quantum key distribution. The idea is to perform QKD from LEO ($\approx 510$ km), with receiver on the ground at an optical ground station. The receiver in the OGS involves homodyning of the signal and the transmitted LO, to extract the information of a specific quadrature from the modulated signal. Additionally, the CV payload also contains another homodyne receiver, that can be used for generation of quantum secure random numbers.

The CV-QRNG module involves homodyning of vacuum states and measuring the output after an amplifier and ADC unit. The sampling rate of the homodyne output is limited by the FPGA to $\approx 500~\text{kHz}$, and the respective outputs from a 12-bit ADC are used to estimate the entropy values as given in Tab. \ref{tab:entropies} and Fig. \ref{fig:entropy}. Note that because of the optical power limitation for the LO for the QNRG, shot noise variance voltage spans only a few bits of the ADC. Thus the limitation in the amount of the extracted random numbers. Higher LO power can reduce the electronic noise with respect to shot noise and also improve voltage span across the ADC for generating larger volume of random numbers. We subject the random numbers from the output to the NIST test suite and the results are given in Fig. \ref{fig:nist}. The NIST test suite involves statistics-based tests, for checking the redundancy and predictability based on a null hypothesis \cite{rukhin2001statistical}.

We also consider the quantum side information leakage to an eavesdropper Eve, and upper bound the $H_{\text{min}}$ as in Ref. \cite{gehring2021homodyne}. Additionally, the noise due to a non-ideal ADC is considered as well. The assumptions made in the manuscript are as follows: the validity of quantum mechanics; homodyne detection of a single mode; the linearity of quadrature measurements; the trustfulness of system characterization; and pure quantum entropy due to vacuum fluctuations being independent of classical correlations.

We have used Toeplitz matrix hashing to extract the random numbers, after the estimation of conditional min-entropy. We have chosen $\epsilon_{\text{hash}} = 10^{-20}$ and this would potentially keep the $\epsilon$-security parameter within the conventional value ($10^{-10}$) \cite{pirandola2020advances}. The dimensionality of Toeplitz matrix is chosen to be 900x200, which yields a 200 bit length of random key for every 100 samples (0.1 ms) from ADC. Since the raw key length is $\approx1$ Mb, we get a total length of $\approx19.5$ Kb of certified random numbers from the 12-bit ADC. Additionally, we have also shown the results for 16-bit ADC, but not from the on-board computer. An important criterion to further explore is taking into account the factor of temperature affecting the various apparatus components and then optimising the conditional min-entropy.

Thus, we provide the results of randomness extraction from the engineering model of the CV payload for SPOQC mission. The extracted random numbers is intended to use for QKD post processing tasks such as basis selection and error correction. Additionally, this also opens up the possibility of using these random numbers to drive the Gaussian distributed modulations for the coherent states. The operational power requirement for CV-QRNG is comparatively lower than single photon based QRNG as there is no need of cooling the photo diodes. Not only that, PIN diode are comparatively radiation tolerant than avalanche photodiode based single photon detectors. These give operational advantage to the CV-QRNG in the constrained space environment. The extracted randomness can also be used for various other applications such as post quantum cryptography that require random numbers in the satellite. Following the successful launch of the SPOQC Cubesat on March 2026, and in-orbit validation, this would constitute the first demonstration of a CV-QRNG operating in space.

\begin{acknowledgments}
V.N.R, K.M, T.P.S and R.K acknowledge the funding support from EPSRC Quantum Communications Hub (Grant number EP/T001011/1) and EPSRC Integrated Quantum Networks Hub (Grant number EP/Z533208/1). E.T.H.M thanks the School of Physics, Engineering and Technology, University of York for PhD funding.
\end{acknowledgments}

\bibliography{references}

\end{document}